\magnification\magstep 1
\baselineskip=0,59 true cm
\vsize=23 true cm
\topinsert\vskip 1 true cm
\endinsert
{\centerline{\bf{$\Gamma(2)$ MODULAR SYMMETRY, RENORMALIZATION}}}
{\centerline{\bf{GROUP FLOW AND THE QUANTUM HALL EFFECT}}}
\vskip 2 true cm
{\centerline{\bf{Yvon Georgelin$^a$, Thierry Masson$^b$ and 
Jean-Christophe Wallet$^a$}}}
\vskip 2 true cm
{\centerline{$^a$Groupe de Physique Th\'eorique, Institut de Physique
Nucl\'eaire}}
{\centerline{F-91406 ORSAY Cedex, France}}
\vskip 1 true cm
{\centerline{$^b$Laboratoire de Physique Th\'eorique (U.M.R. 8627), 
Universit\'e de Paris-Sud}}
{\centerline{B\^at. 211, F-91405 ORSAY Cedex, France}}
\vskip 2 true cm
{\bf{Abstract:}} We construct a family of holomorphic $\beta$-functions whose RG flow
preserves the $\Gamma(2)$ modular symmetry and reproduces the
observed stability of the Hall plateaus. The semi-circle law relating the
longitudinal and Hall conductivities that has been observed experimentally
is obtained from the integration of the RG equations for any permitted transition 
which can be identified from the selection rules encoded in the flow diagram. 
The generic scale dependance of the conductivities is found to
agree qualitatively with the present experimental data. The existence of a
crossing point occuring in the crossover of the permitted transitions is
discussed. 
\vskip 4 true cm
(May 1999)
\vskip 0.5 true cm
LPT-99/02 
\vfill\eject
{\noindent{{\bf{1. INTRODUCTION}}}}\par
\vskip 0,5 true cm
The Quantum Hall Effect (QHE) is a remarquable phenomenon occuring in a
two-dimensional electron gas in a strong magnetic field at low temperature [1].
Since the discovery of the quantized integer [2] and fractional [3] Hall
conductivity, the QHE has been an intensive field of theoretical and
experimental investigations. The pioneering theoretical contributions [4]
analyzing the basic features of the hierarchy of the Hall plateaus
have triggered numerous works aiming to provide a better understanding of
the underlying properties governing the complicated phase diagram associated
with the quantum Hall regime together with the precise nature of the various
observed transitions between plateaus and/or focusing on a 
characterization of a suitable theory.\par
It has been realized for some time that modular symmetries may well be of
interest to understand more deeply salient properties of the QHE. For
instance, the superuniversality proposed in [5] to explain the apparent
similarity of the observed transitions is reminiscent to modular
tranformations. Besides, it has been shown that some properties of the
phase diagram may well be explained in terms of modular group
transformations in a two-parameter scaling theory. At the present time, a
fully satisfactory microscopic or effective theory for the QHE, from which
the relevant modular symmetry (if any) would come out, is still lacking.
This has somehow motivated studies focalized on the derivation of general
constraints on the phase diagram (and/or expressions for the conductivities)
coming from the full modular group [6] or some of its subgroups [7]. Indeed,
it is well known that the existence of a discrete symmetry group acting on
the parameter space of a theory induces restrictions on the renormalization
group (RG) flow. This has been pointed out [6],[7] in the case of the full
modular group which in that context can be viewed as a rich extension of the
old Kramers-Wannier $Z_2$ duality of the two-dimensional Ising model. This
interesting aspect has been applied in various areas of physics, such as 
statistical systems [8], extended Sine-Gordon theories [9], as well as
non-perturbative analysis of $N=2$ supersymmetric Yang-Mills theory [10].\par
In this paper, we construct a familly of holomorphic $\beta$-functions 
which reproduces the observed stability of the Hall plateaus and whose 
corresponding RG flow in the conductivity plane (i.e. the parameter space) 
preserves a $\Gamma(2)$ symmetry acting on it. The paper is 
organized as follows. The section 2 is devoted to the
construction. In section 3, we discuss the corresponding 
physical consequences. We show in particular that the recently 
observed semi-circle law [11] relating the longitudinal and Hall 
conductivities can be recovered from the integration
of the RG equations and that the predicted crossover for the various
transitions is found to be in good qualitative agreement with the present
experimental observations. We also compare our results to those obtained
in a recent work dealing with the construction of a $\beta$-function based
on another (larger) subgroup of the modular group [7c]. In section 4, we
collect the main results of this paper and we conclude.\par
\vskip 1 true cm
{\noindent{{\bf{2. CONSTRUCTION OF THE $\beta$-FUNCTIONS}}}}\par
\vskip 1 true cm
{\noindent{{\bf{2.1 Basic properties of $\Gamma(2)$}}}}\par
\vskip 0,5 true cm
The properties of the modular group $\Gamma(1)$($\equiv\ PSL(2,Z)$)
and its various subgroups can be found
in [12]. In this section we collect all the relevant ingredients that we will
use in the subsequent analysis. Let ${\cal{P}}$ and $z$ denote respectively 
the open upper-half complex plane and a complex coordinate on
${\cal{P}}$ (Im$z>0$). One defines ${\bar{\cal{P}}}$$\equiv{\cal{P}}\cup Q$,
where $Q$ is the set of rational numbers. We first recall that the group $\Gamma(2)$ is the
set of transformations $G$ acting on ${\bar{\cal{P}}}$ defined by:
$$G(z)={{az+b}\over{cz+d}},\ \ \ a,b,c,d\in Z,\ \ 
(a,d)\ {\hbox{odd and}},\ (b,c)\ {\hbox{even}} \eqno(2.1a)$$ 
$$ad-bc=1\ \ ({\hbox{unimodularity condition}})  \eqno(2.1b).$$
$\Gamma(2)(\subset\Gamma(1))$, is the free group generated by
$$T^2(z)=z+2,\ \ \Sigma(z)=ST^{-2}S(z)={{z}\over{2z+1}} \eqno(2.2),$$
where $T(z)=z+1$ and $S(z)=-{{1}\over{z}}$ are the two generators of 
the modular group $\Gamma(1)$ and is known in the mathematical litterature 
as the principal congruence unimodular group at level 2. The 
corresponding principal fundamental domain ${\cal{D}}_{\Gamma(2)}$, depicted 
on fig.1, has three cusps denoted
by [0], [1]$\simeq$[$-1$] and [$i\infty$] that are respectively identified
with the three points 0, 1 and $i\infty$ of ${\bar{\cal{P}}}$ which are the only
fixed points of $\Gamma(2)$ on ${\cal{D}}_{\Gamma(2)}$. The whole set 
of fixed points of $\Gamma(2)$ on ${\bar{\cal{P}}}$ is obtained as usual 
by successive $\Gamma(2)$ transformations of these three
points{\footnote\ddag{\sevenrm{Observe that $\Gamma(2)$ has only real fixed
points.}}}. Notice the identification of the frontiers 
on ${\cal{D}}_{\Gamma(2)}$ as indicated on fig.1.  \par
It has been pointed out recently in [7b,13] that $\Gamma(2)$
can be used to derive a model for a classification of fractional (as well as
integer) Hall states. This classification, which refines the Jain one [14] 
and involves a kind of generalization of the "law of the corresponding states"
derived in [5], seems to reproduce successfully the observed hierarchical
structure of the Hall states. The salient feature of
the proposed construction is that each family of quantum fluid states
indexed by fractions with odd denominators (plus the insulator state(s)) is
generated from a metallic state labelled by an even denominator fraction
through specific $\Gamma(2)$ transformations. To be more precise, we first
rewrite the transformations (2.1a,b) as
$$G(z)={{(2s+1)z+2n}\over{2rz+(2k+1)}},\ \ (2s+1)(2k+1)-4rn=1  \eqno(2.3a,b)$$
where $k,\ n,\ r,\ s\in Z$. We identify for the moment $z$
with a filling factor $\nu=p/q$ and select a given Hall metallic state
labelled by $\lambda={{(2s+1)}\over{2r}}$ ($r\ge0,s\ge0$). Then, as shown in
[13], one obtains a hierarchy of Hall (liquid) states surrounding the metallic state $\lambda$
from the images $G^\lambda_{n,k}(0)$ and $G^\lambda_{n,k}(1)$ of 0
and 1 by the family of transformations $G^\lambda_{n,k}\in\Gamma(2)$ ($n$
and $k$ satisfying (2.3b)) which sends $z=i\infty$ onto $\lambda$. As an
exemple, the double Jain family of states surrounding the metallic state
$\lambda=1/2$ 
$${{1}\over{3}},\ {{2}\over{5}},\ {{3}\over{7}},\dots,\ {{N}\over{2N+1}}
\eqno(2.4a),$$
$${{2}\over{3}},\ {{3}\over{5}},\ {{4}\over{7}},\dots,\ {{N}\over{2N-1}}
\eqno(2.4b),$$
can be easily recovered in this scheme from the images
$G^{1/2}_{n,k}(0)$ and $G^{1/2}_{n,k}(1)$ with respectively $n\ge0$ for (2.4a)
and $n<0$ for (2.4b). We recall that this construction separates the even
numerator Hall fractions from the odd numerator ones so that it may be
possible to take into account a possible particle-hole symmetry within the
present scheme. Other families surrounding any state indexed by an
even denominator (metallic state) can be constructed in the same way so that
all the experimentally observed Hall states can be taken into account in the
present construction. It has been further shown in [7b]
that the corresponding predicted global organization of the various Hall
conductivity states stemming from the action of $\Gamma(2)$ fits quite well
with (some of) the present experimental data.\par
The possible important role played by modular symmetries in the QHE has been
considered for some time. Some of the related works have emphasized that
(most of) the main features of the (up to now) experimentaly observed phase
structure of QHE seem to be recovered from the action of suitable subgroup
of the modular group on the complex conductivity plane hereafter identified
with
${\bar{\cal{P}}}$ and parametrized by
$z={{\hbar}\over{e^2}}(\sigma_{xy}+i\sigma_{xx})$
{\footnote\ddag{\sevenrm{in the following, $e^2$=$\hbar=1$}}} (with
Im$z\equiv\sigma_{xx}$$\ge0$). In particular the group $\Gamma_0(2)$ has
been mostly considered by some authors and used recently in [7c] to
constraint the $\beta$-function governing the Renormalization Group (RG) flow
of the conductivities for Quantum Hall systems. The corresponding studies
have been performed under various physically acceptable 
set of hypothesis. In the 
next subsection, we will consider somehow similar hypothesis to
study the restrictions which can be obtained from $\Gamma(2)$ on the
$\beta$-function of the RG flow for the conductivities. We therefore assume
that the action of $\Gamma(2)$ on real filling factors $\nu=p/q$ that has
been described above can be extended to an action on ${\bar{\cal{P}}}$.\par 
\vskip 1 true cm
{\noindent{{\bf{2.2 Holomorphic $\beta$-functions from $\Gamma(2)$ symmetry}}}}\par
\vskip 0,5 true cm
Let $t$ be a scale parameter whose possible explicit form
will be specified in section 3. We first recall that scale transformations on the
complex conductivity plane ${\bar{\cal{P}}}$ generate a 
RG flow $z\to R(t;z,\bar z)\equiv z(t)$, from
which the $\beta$-function is defined to be the (contravariant) vector field
tangent to this flow, namely: 
$$\beta(z,\bar z)={{dR(t;z,\bar z)}\over{dt}}={{dz(t)}\over{dt}} \eqno(2.5).$$
It is well known that the existence of a discrete symmetry group acting on the
parameter space of a theory may induce restrictions on the RG flow and, in
turn, provides some non-perturbative information on the RG flow (stemming
basically from reasonable ans\"atze for the corresponding
$\beta$-functions). As already mentionned in the introduction, this aspect 
has already been investigated in
various areas of physics. Most of the considerations 
involved in these corresponding 
works can be adapted to the present situation for which we now outline the
main steps of the analysis.\par
First of all, the crucial mathematical hypothesis is that the action of
$\Gamma(2)$ commutes with the RG flow, which basically means that if the
$\Gamma(2)$ symmetry of the parameter space (that is in the present case the
conductivity plane) holds at a given scale, it will be preserved by the RG
downwards to lower scales. This hypothesis in particular determines the
$\Gamma(2)$-transformations of the $\beta$-function, given by
$$\beta(G(z),{\overline{ {G(z)}}})=(cz+d)^{-2}\beta(z,\bar z) \eqno(2.6),$$
for any $G\in\Gamma(2)$, and may account for the apparent observed
superuniversality in the Quantum Hall transitions
[5]{\footnote*{\sevenrm{It is easy to show that
distinct critical points of the RG flow related by a $\Gamma(2)$
transformation will have the same scaling exponents.}}}.\par
Eqn. (2.6) indicates that $\beta$ transforms as a modular form of
$\Gamma(2)$ with weight $-2$ whenever $\beta$ is {\it{holomorphic in}} $z$ on
${\cal{P}}$. In this later case, the applications of general results
stemming from complex analysis permits one already to constraint strongly the
possible expression for an admissible $\beta$-function. In the rest of this
section, we will therefore assume that $\beta$ is holomorphic in $z$. This
hypothesis will be commented upon in the beginning of section 3. \par
Now, a general theorem on modular forms [12] states that
any modular form $\omega(z)$ of any subgroup $\Gamma$ (of finite 
index) of the modular group with even
weight $k$ can be represented on ${\cal{D}}_{\Gamma}$, the fundamental domain
of $\Gamma$, as
$$\omega(z)=(\lambda^\prime(z))^{k/2}R(\lambda) \eqno(2.7),$$
where $\lambda(z)$ is a
modular function of $\Gamma${\footnote\dag{\sevenrm{that is, a function
invariant under the action of $\Gamma$}}} 
defined on ${\cal{D}}_{\Gamma}$, $\lambda^\prime(z)={{d\lambda(z)}\over{dz}}$ and
$R(\lambda)$ is a rational function in $\lambda$. In the present case,
$k=-2$ and $\lambda$, the modular function of $\Gamma(2)$, can be chosen as
$$\lambda={{\theta^4_2}\over{\theta^4_3}}  \eqno(2.8),$$
and satisfies on ${\cal{D}}_{\Gamma(2)}$
$$\lambda(i\infty)=0,\ \lambda(0)=1,\ \lambda(1)=\infty  \eqno(2.9).$$
In (2.8), the Jacobi $\theta$ functions $\theta_2$ and $\theta_3$
(together with $\theta_4$ given here for further convenience) are defined by
$$\theta_2=2\sum_{n=0}^{\infty}q^{(n+{{1}\over{2}})^2}=2q^{{1}\over{4}}
\prod_{n=1}^\infty(1-q^{2n})(1+q^{2n})^2\eqno(2.10a),$$
$$\theta_3=\sum_{n=-\infty}^{\infty}q^{n^2}=\prod_{n=1}^\infty(1-q^{2n})(1+q^{2n-1})^2\eqno(2.10b),$$
$$\theta_4=\sum_{n=-\infty}^{\infty}(-1)^n
q^{n^2}=\prod_{n=1}^\infty(1-q^{2n})(1-q^{2n-1})^2\eqno(2.10c),$$
where $q=\exp(i\pi z)$.\par
Therefore, consistency with the previous two hypothesis requires the
general expression for the $\beta$-function on ${\cal{D}}_{\Gamma(2)}$ to be
$\beta(z)=\lambda^{\prime}(z)^{-1}R(\lambda)$ with $\lambda$ given in (2.8), which
can then be straightforwardly extended to ${\cal{P}}$ by the action of
$\Gamma(2)$. We have now to 
confront this general expression to the experimental situation.\par 
The strongest experimental constraint comes from the observed stability of
the Hall plateaus labelled by integer as well as odd denominator fractional
filling factors, which must presumably correspond to attractive stable fixed
points of the $\beta$-function. To apply this constraint, we proceed as
follows. First, observe that in the present framework, the fixed points of
$\Gamma(2)$ must be by construction critical points of the $\beta$-function.
On ${\cal{D}}_{\Gamma(2)}$, the only fixed points of $\Gamma(2)$ are 0,
1 and $i\infty$ which can be respectively identified to the (Hall) insulator
state, the first Landau level and some (unobserved) superconducting state.
Next, observe that in the classification of the Hall states based on the
$\Gamma(2)$ symmetry [7b,13], the Hall plateaus correspond to the images of 0 and 1
by $\Gamma(2)$ as recalled in section 2.1 (the images of $i\infty$
correspond to even denominator (metallic) Hall states). From 
these observations and under the
further assumption that $\beta(z)$ has no other critical points than those
given by the fixed points of $\Gamma(2)$, it is easy to realize that
$\beta(z)$ can be conveniently parametrized on ${\cal{D}}_{\Gamma(2)}$ as
$$\beta(z)={{\alpha}\over{\lambda^\prime(z)}}\lambda^p(z)(\lambda(z)-1)^q
\eqno(2.11),$$
where $\alpha$ is a complex 
constant, $\lambda$ is still given by (2.8), $p,q\in Z$ and use has been 
made of (2.9).\par
Now, according to the previous discussion, eqn.(2.11) must have zeros at
$z=0$ and $z=1$. This is realized provided
$$q\ge 1  \eqno(2.12a),$$
for $\beta(z=0)=0$ and
$$p+q-1\le 0  \eqno(2.12b),$$
for $\beta(z=1)=0$. These constraints can be easily derived by combining
(2.11) and (2.9) with the explicit expression for $\lambda^\prime(z)$ given
by $\lambda^\prime(z)=i\pi\lambda(z)\theta^4_4(z)$, obtained from (2.8) and 
the functional relation 
$${{\theta^\prime_2}\over{\theta_2}}-{{\theta^\prime_3}\over{\theta_3}}={{i\pi}\over{4}}
\theta^4_4  \eqno(2.13),$$
and making use of the following asymptotic expansions for the Jacobi $\theta$ functions:
$$\theta_2(z)\sim\sqrt{{{i}\over{z}}},\ \ \theta_3(z)\sim\sqrt{{{i}\over{z}}}
,\ \ \theta_4(z)\sim\sqrt{{{i}\over{z}}}\exp(-{{i\pi}\over{4z}})\ \ \
{\hbox{for}}\ z\sim 0  \eqno(2.14a),$$ 
$$\theta_2(z)\sim\sqrt{{{i}\over{z-1}}},\ \
\theta_3(z)\sim\sqrt{{{i}\over{z-1}}}\exp({-{i\pi}\over{4(z-1)}})
,\ \ \theta_4(z)\sim\sqrt{{{i}\over{z-1}}}\ \ \
{\hbox{for}}\ z\sim 1  \eqno(2.14b).$$
>From the combination of (2.11) with (2.12a) and (2.12b), one easily realize
that $\beta(z)$ must be singular for $z=i\infty$, using for instance
$$\theta_2(z)\sim\exp({{i\pi z}\over{4}}),\ \ \theta_3(z)\sim 1,\ \
\theta_4(z)\sim 1\ \ {\hbox{for}}\ z\to i\infty  \eqno(2.15),$$
a result that can be expected from general results on holomorphic
functions. Finally, we find that the complex constant $\alpha$ appearing in
(2.11) must be chosen to be real in order to obtain from the $\beta$-function a
flow with the desired properties, in particular involving stable fixed
points at values corresponding to the Hall plateaus. This is illustrated
on fig. 2 and 3 where the flow of (2.11) with $q=1,\ p=0$ and $\alpha=-1$ is
represented.\par
\vskip 1 true cm
{\noindent{{\bf{3. DISCUSSION}}}}\par
\vskip 1 true cm
Let us first summarize what has been derived up to now.
Putting (2.8), (2.11), (2.12) together with
$\lambda^\prime(z)=i\pi\lambda(z)\theta^4_4(z)$, we finally obtain
$$\beta=i^{2q-1}{{\alpha}\over{\pi}}{{\theta_2^{4(p-1)}\theta_4^{4(q-1)}}\over{\theta_3^{4(p+q-1)}}}$$
$$p,q\in Z,\ q\ge1,\ p+q-1\le0,\ \alpha<0 \eqno(3.1),$$
which is defined on ${\cal{D}}_{\Gamma(2)}$ and can be straightforwardly
extended to ${\bar{\cal{P}}}$. This is the main result of the 
previous section. Eqn (3.1) represents a
physically admissible familly of {\it{holomorphic}} $\beta$-functions
reproducing in particular the experimentally observed stability of the Hall
plateaus and whose corresponding RG flow in the complex conductivity plane
(i.e. the parameter space) preserves a $\Gamma(2)$ symmetry acting on it.\par
In this section we will show that the physical predictions that can be
extracted from (3.1) are in good agreement with the present experimental
observations. Namely, we will show that the "semi-circle"law which has been
recently observed [15] can be recovered from the behaviour of $\sigma_{xy}$
and $\sigma_{xx}$ obtained from the integration of (3.1). We will also show
that the crossover in the plateau-plateau and plateau-insulator (observed)
transitions described by (3.1) agrees qualitatively with the present
experimental observations.\par
Before starting the discussion, some important remarks concerning the
holomorphy hypothesis as well as the already proposed candidates for
$\beta$-functions are in order. As it is well known, the 
holomorphy constraint is a very strong one which
restricts severely the possible expression for $\beta$. Relaxing this
constraint permits one to have much more freedom for the construction of
physically admissible $\beta$-functions. For a recent comprehensive analysis
of the non-holomorphic case, see e.g. [16] and references therein. Here, we
notice that non-holomorphic $\beta$ that have been proposed in [16] can
obviously be related through their asympotic (large $\sigma_{xx}$) behaviour
to the $\beta$-functions stemming from the dilute-instanton gas calculations
performed in the framework of non-linear sigma-models [17], whereas the family
of holomorphic $\beta$-functions given in (3.1) cannot be. In particular,
the corresponding asymptotic (large $\sigma_{xx}$) behaviour is different
and indeed cannot be finite as it is for those non-holomorphic $\beta$'s.
Imposing finitude at large $\sigma_{xx}$ for (3.1) would produce necessarely
unstable Hall plateaus, in clear contradiction with the experimental
observations. Therefore, one of the characteristic features of (3.1) is the
existence of a singularity as $\sigma_{xx}\to\infty$ and consequently at any
even denominator rational point on the real axis in the conductivity
plane. Anticipating the discussion, the physical consequence on the
corresponding flow diagram is the existence of unstable path connecting even
denominator (metallic) Hall states to odd denominator (or insulator) states
which might be associated to "unfavored" (but nevertheless observable) transitions.\par
Let us now turn to already proposed candidates for a $\beta$-function and
RG flows. Recall that the early developments on this subject were essentially
based on a field theory of the type of non linear sigma model [17] (see also
[18]) mentionned just above with two (dimensionless) coupling constants identified with the
longitudinal ($\sigma_{xx}$) and Hall ($\sigma_{xy}$) conductivities of the
disorderd electron gas. Starting from this framework, a RG flow diagram for
the conductivities has been conjectured [18] whose characteristic feature is the
existence of fixed points occuring at some $\sigma_{xx}$ and $\sigma_{xy}$
equal to half integer values. Althought this proposal is appealing and seems
to capture some experimental features of the (mainly integer) Quantum Hall
Effect, it is plagued with a problem. Indeed, the postulated fixed points
(if they exist as fixed points of the non linear sigma model [17], a fact
which is not clear at the present time) correspond to a small $\sigma_{xx}$
(strong coupling) regime whereas the dilute-instanton gas calculation
that gives rise to the non-holomorphic $\beta$-function underlying the
conjectured RG flow is valid only in the large $\sigma_{xx}$ (weak coupling)
regime. Whether this crude extrapolation from the weak to the strong
coupling regime is finally correct or not is unclear at the present time.
Keeping in mind the above remarks and that there is no experimental facts
favoring either holomorphic or non-holomorphic $\beta$ as far as we know,
one can reasonably regard the family of holomorphic $\beta$'s (3.1) based on
$\Gamma(2)$ as possible candidates for a description of aspects of the
physics of the QHE effect.\par
We close these remarks by noticing that
similar constructions of holomorphic $\beta$-functions
based on larger subgroups of the full modular group have been performed
recently, most of them focussing on the subgroup $\Gamma_0(2)$. Recall that
$\Gamma_0(2)$ is generated by $T(z)=z+1$ and $\Sigma(z)=z/(2z+1)$ and that one has
$\Gamma(2)\subset\Gamma_0(2)$, i.e. $\Gamma(2)$ is a subgroup of
$\Gamma_0(2)$. It has been shown [7c] that a $\Gamma_0(2)$-based construction
gives rise to a qualitative behaviour of the crossover in the various transitions 
that is similar to the one obtained from our $\Gamma(2)$-based construction.
There are however specific differences appearing in the corresponding flows. In
particular, one of the fixed point of $\Gamma_0(2)$ in its fundamental domain
has to be identified naturally with the crossing point appearing in the
plateau-insulator transitions, whose position is therefore entirely
("rigidely") determined in the conductivity plane. Recall that it corresponds to $\sigma_{xy}=1/2$ and
$\sigma_{xx}=1/2$ ($z={{1+i}\over{2}}$) for the $0\to1$ transition.
Moreover, the proposed $\beta$-function has a pole at this point. In that
case, the unique path in the conductivity plane connecting $z=0$ to $z=1$,
which must go obviously through the pole $z={{1+i}\over{2}}$, appears to be
unstable (and indeed can evolve either from $0$ toward $\infty$ or to $0$
toward $1/2$), a fact that can be easily verified numerically by plotting
the flow generated by the corresponding $\beta$. Further investigations are
needed to clarify the experimental and theoretical status of this predicted
crossing point. In the $\Gamma(2)$ case, the situation is quite different.
Indeed, the point $z={{1+i}\over{2}}$ does not play a distinguished role
simply because it is not a fixed point of $\Gamma(2)$ so that consequently
it can be neither a zero nor a pole of the corresponding $\beta$ in the
present framework.\par
Let us now examine critically the physical consequences encoded in (3.1).
First, notice that $\beta$-functions defined by (3.1) and such that
$p+2q-2=0$ holds generate a flow approaching $z=0$ and $z=1$ in the same
manner. This can be easily seen by combining (3.1) with (2.14a,b) and
studying the behaviour of $\beta$ in the vicinity of $z=0$ and $z=1$. In
particular, it can be straightforwardly realized that the fastest approach of
0 and 1 is obtained when $q=1$ and $p=0$ which corresponds to asymptotic
expressions for $\beta$ that do not involve exponential factors. From now on
we will focuss on this later situation.\par
Now, we point out that the RG equation can be formally integrated along
the single trajectories in the parameter space, leading
to an algebraic relation between $\sigma_{xy}$ and $\sigma_{xx}$ which
reproduces the "semi-circle" law that has been experimentaly observed, at
least at low temperature, in the
study of the plateau-insulator and plateau-plateau transitions [15]. Indeed,
combining (2.5) with (2.11) (in which now $p=0$, $q=1$
and we take $\alpha=-1${\footnote\dag{\sevenrm{other real value for $\alpha$
corresponds to simple rescaling of $t$ so that the ensuing analyzis is not
modified}}}), one obtains
$$dt=-{{d\lambda}\over{\lambda-1}}  \eqno(3.2),$$
where $\lambda$ is still given by (2.8). The integration of (3.2) gives
$$t=-log(\lambda-1)+\chi  \eqno(3.3),$$
where $\chi$ is a complex constant and $log$ denotes a determination of the
complex logarithm. It follows, for any determination of the
logarithm, that
$$\lambda=1+\exp(\chi-t) \eqno(3.4),$$
and one has $\lambda\to 1$(resp.$+\infty$) for $t\to+\infty$
(resp.$-\infty$).\par
Let us assume $\chi$ real so that $\lambda$ is real for any $t$. Then, it is a general
result that the inversion of the map
$z(t)\mapsto\lambda(z(t))=1+\exp(\chi-t)$, with $\lambda$ given by (2.8), 
gives rise to a curve $t\mapsto z(t)$ which is a semi circle linking two rational numbers 
as end-points on the real axis
together with a point $z_0$ located on this semi-circle and satisfying $\lambda(z_0)=1+\exp\chi$. 
The inversion of this map does not result in a unique semi-circle because if
$t\mapsto z(t)$ is a solution then, for
any $G\in\Gamma(2)$, $t\mapsto G(z(t))$ is also a solution
(whose end-points on the real axis are therefore the images by $G\in\Gamma(2)$
of the end-points of the initial solution). But it can be easily realized that 
the choice of such a solution among all the possible one is completely
equivalent to the choice of a $z_0$ such that $\lambda(z_0)=1+\exp\chi$,
which can be regarded as initial conditions for the RG equation.\par
It is instructive to attempt to make a closer contact with the experimental
situation throught the following (more physical) interpretation for $z_0$.
Observe first that one should have $t=0$ at $z=z_0$. Assume then that $t$
can be cast into the form [19]
$$t=f({{1}\over{T^\gamma}}({{1}\over{B}}-{{1}\over{B_c}})^\delta)\eqno(3.5),$$
at least in the domain of interest, a form which is inspirated from a
two-parameter scaling
framework, where $f$ is a monotonic function with $f(0)=0$, $B_c$ is a critical value 
for the external magnetic field
$B$ and $T$ is the temperature. This therefore suggests that $z_0$ would 
correspond to the point on the semi-circle where $B=B_c$, independantly of
the temperature, so that $z_0$ might be naturally identified with a crossing point
appearing in the crossover between the two Hall states whose filling factors
correspond to the rational points on the real axis connected by the
semi-circle.\par
To illustrate the previous discussion, a representative exemple of 
the resulting flow is depicted on figure 2 where only semi-circle line flows
are considered. The plateau-insulator transition
$0\to1$ corresponds to the (full line) large semi-circle linking $0$ and
$1$. As it should be clear now, the action of successive 
$\Gamma(2)$ transformations on this semi-circle
(which may be viewed as a template for the transitions) gives rise to the
other (full-line) semi-circles appearing on this figure. For instance, the
following transformation $G(z)={{3z-2}\over{2z-1}}$$\in\Gamma(2)$ maps the
semi-circle for the $0\to1$ transition into the semi-circle connecting $1$ and $2$,
which therefore corresponds to a direct transition between the Hall plateaus
with integer filling factor $\nu=1$ and $\nu=2$, whereas
$G(z)={{z}\over{2z+1}}$$\in\Gamma(2)$ maps the $0\to1$ semi-circle into the one
associated to the transitions $0\to1/3$. Therefore, if the present framework
is correct, it is expected that such a semi-circle law should be
experimentally observed for any other permitted transitions. This is already
the case for the observed $0\to1/3$, $0\to1$ and $1\to2$ transitions that
have been recently studied in e.g. some Si MOSFET devices in the Quantum
Hall regime [20,15b] and for which the data on $\sigma_{xx}$ and $\sigma_{xy}$
fit well with the semi-circle law expected from the present
$\Gamma(2)$-based framework. Notice by the way that selection rules for the
permitted direct plateau-plateau and plateau-insulator transitions can be
obviously extracted from the flow diagram depicted on fig.2 by simply
observing that each permitted transition is rigidely linked with a semi-circle
whose end-points on the real axis correspond to the filling factors
labelling the transition.\par
The above analysis can be easily adapted to the case where $\chi$ is a
complex number. In this later situation, the trajectories 
$t\mapsto z(t)$ (each of which still connecting two rational numbers on the real axis) 
are no longer semi-circles, as depicted on fig.3.\par
As announced in the beginning of this section, there are a priori possible
transitions of the type insulator$\to$ even denominator (metallic) state as
well as odd denominator $\to$ metallic state, namely $0\to1/2$, $1\to1/2$
and the corresponding images by $\Gamma(2)$. Some representative examples of
these transitions are indicated on fig.2 and 3 by dashed lines 
{\footnote\dag{\sevenrm{Notice that there exists also, as a consequence
of the present construction, transitions of the type $0\to\infty$ or
$1\to\infty$ and the corresponding $\Gamma(2)$ images. The state
$\sigma_{xx}=\infty$ would represent some (yet unobserved) supraconducting
state.}}}. The associated trajectories are in fact unstable as any small
deviation from the semi-circle connecting, says $0$ to $1/2$ will give rise
to a quite different transition. For instance, it can be easily realised
from fig.3 that a small perturbation of the flow line associated to the
$0\to1/2$ transition will give rise to either the $0\to1$ or $0\to1/3$
transition, whereas any plateau-plateau as well as plateau-insulator
transition is stable against perturbation.\par
The explicit expressions for $\sigma_{xx}$ and $\sigma_{xy}$ as
functions of $t$, which are expected to provide the qualitative behaviour of
the crossover for a transition, can be easily extracted from (3.2)-(3.4).
The method is standard and is essentially similar to the one used in [7c]. The
resulting expressions are similar to the one given in [7c]. It is however
interesting to plot them numerically. This is done on fig.4 (resp. fig.6)
for the $1\to0$ (resp. $1\to2$) transition, whereas fig.5 and fig.7
represent the $t$-dependance of the corresponding resistivities. It can be
realized that the predicted behaviour of the conductivities (as well as the
resistivities) is in good qualitative agreement with the experimental 
one reported in particular in [15a,b]. We observe by the way that the "almost
linear" asymptotic shape of $\rho_{xx}(t)$ for $t>0$ that we obtain fits well with
a recent experimental measurement of the corresponding quantity for the
$1\to0$ transition [15b] (keeping in mind the possible interpretation for $t$
parametrized by (3.5)). Notice that somehow similar results for the
crossovers have been obtained in [7c] where the symmetry $\Gamma_0(2)$ is
considered instead of $\Gamma(2)$.\par
As a last remark, notice that a precise experimental determination of 
the location of the crossing point on the semi-circular trajectories in the
conductivity plane would be helpful for the identification of the possibly
relevant modular subgroup. Indeed, a crossing point found to be really
located on the uppermost point of the semi-circle (e.g. corresponding to $z={{1+i}\over{2}}$ 
for the $0\to1$ transition) would favor the group $\Gamma_0(2)$. On the
other hand, a crossing point found to deviate significantly from the
uppermost location would favor $\Gamma(2)$ (putting therefore $\Gamma_0(2)$
into trouble). This later case seems to have been recently observed in the
experiment performed in the second ref.[15b]. Note that, in this experiment, the corresponding candidates for 
the crossing points in the $0\to1$ and $0\to 1/3$ transitions can be related
to each other (within, says, $\sim$ $10\%$) by a $\Gamma(2)$
transformation, as it must be the case in the present framework.\par
\vskip 1 true cm
{\noindent{{\bf{4. CONCLUSION}}}}\par
\vskip 1 true cm
Let us summarize in physical words the main results of this paper. We have
proposed a (familly of) holomorphic $\beta$-function(s) whose RG flow
preserves the $\Gamma(2)$ modular symmetry and which is consistant with the
observed stability of the Hall plateaus. The semi-circle law relating the
longitudinal and Hall conductivities that has been observed experimentally
for the $0\to1$, $0\to1/3$ and $1\to2$ transitions is obtained from the
integration of the RG equations for these transitions and, in fact, must 
hold in the present framework for any permitted transition which can be 
easily identified from the selection rules encoded
in the flow diagram. Moreover, it has been shown that there exists a unique
point on each semi-circle where the generic scale parameter $t$ vanishes.
This combined with a two-parameter scaling hypothesis, as an additional
phenomenological input (yielding a plausible parametrization for $t$ given
by (3.5)), suggests to interpret this point as the crossing point occuring
in the crossover of the two Hall states involved in the transition.The 
generic scale dependance of the conductivities has been verified to
agree qualitatively with the present experimental data. In the present
framework, the trajectories in the conductivity plane involving an even
denominator filling factor are found to be unstable. Although we do not have
a clear interpretation (if any) of this, one might expect that the
corresponding (observed) transitions are unfavored.\par
\vfill\eject
{\noindent{\bf{REFERENCES}}}
\vskip 1 true cm
\item{1)} The Quantum Hall Effect, 2nd ed., R.E. Prange and S.M. Girvin eds.
(Springer-Verlag, New-York) 1990. See also Perspectives in Quantum Hall
Effect, S.D. Sarma and A. Pinczuk eds. (Wiley, New York) 1997.\par
\item{2)} K.von Klitzing, G. Dorda and M. Pepper, Phys. Rev. Lett. 45 (1980)
494.\par 
\item{3)} D.C. Tsui, H.L. St\"ormer and A.C. Gossard, Phys. Rev. Lett. 48
(1982) 1559.\par
\item{4)} R.B. Laughlin, Phys. Rev. Lett. 50 (1983) 1385; F.D. Haldane,
Phys. Rev. Lett. 51 (1983) 605; B.I. Halperin, Phys. Rev. Lett. 52 (1984)
1583.\par
\item{5)} S. Kivelson, D.H. Lee and S.C. Zhang, Phys. Rev. B46 (1992)
2223.\par 
\item{6)} C.A. L\"utken and G.G. Ross, Phys.Rev. B45 (1992) 11837, Phys. 
Rev. B48 (1993) 2500; see also 
E. Fradkin and S. Kivelson, Nucl. Phys. B474 (1996) 543.\par 
\item{7a)} C.A. L\"utken, Nucl. Phys. B396 (1993) 670.\par 
\item{7b)} Y. Georgelin, T. Masson and J.C. Wallet, J. Phys. A: Math. Gen. 30 
(1997) 5065.\par 
\item{7c)} B.P. Dolan, Modular invariance, Universality and crossover in the 
QHE, cond-mat/9809294;
          B.P. Dolan, Duality and the modular group in QHE,
	  cond-mat/9805171.\par
\item{8)} J. Cardy, Nucl. Phys. B203 (1982) 17; J. Cardy and E. Rabinovici,
Nucl. Phys. B203 (1982) 1.\par 
\item{9)} D. Carpentier, J. Phys. A: Math.Gen.32 (1999) 3865 and references therein.\par  
\item{10)} N. Seiberg and E. Witten, Nucl. Phys. B426 (1994) 19; J.I.
Latorre and C.A. L\"utken, Phys. Lett. B421 (1998) 217; A. Ritz, Phys. Lett.
B434 (1998) 54.\par 
\item{11)} A.M. Dykhne and I.M. Ruzin, Phys. Rev. B50 (1994) 2369; I.M.
Ruzin and S. Feng, Phys. Rev. Lett. 74 (1995) 154.\par  
\item{12)} R.A. Rankin, Modular Forms and Functions, Cambridge University
Press, 1977.\par
\item{13)} Y. Georgelin and J.C. Wallet, Phys. Lett. A224 (1997) 303.\par
\item{14)} J.K. Jain, Phys. Rev. B41 (1990) 7653.\par
\item{15a)} M. Hilke et al., Nature 395 (1998) 675.\par 

\item{15b)}  M. Hilke et al., Semi circle: An exact relation in the Integer or
Fractional Quantum Hall Effect, cond-mat/9810217.\par
\item{16)} C.P. Burgess and C.A. L\"utken, On the implication of discrete
symmetries for the $\beta$-function of Quantum Hall System,
cond-mat/9812396.\par
\item{17)} A.M.M. Pruisken quoted in ref. [1a], Phys. Rev. B32 (1985) 2636,
Nucl. Phys. B285 (1987) 61.\par

\item{18)} D.E. Khmel'niskii, Pisma Zh. Eksp. Teor. Fiz.38 (1983) 454.\par 

\item{19)} For a general discussion see B. Huckestein, Rev. Mod. Phys. 67
(1995) 357.\par

\item{20)} D. Shahar et al., Phys. Rev. Lett. 79 (1997) 479.\par  
\end